\documentclass[twocolumn,prl,floatfix,superscriptaddress]{revtex4}
\usepackage{graphicx}
\usepackage{latexsym}
\usepackage{amssymb}

\begin{document}

\title{Anomalous magnetoresistance peak in (110) GaAs two-dimensional holes: 
\\ Evidence for Landau-level spin-index anticrossings}

\author{F. Fischer}
\affiliation{Walter Schottky Institut, Technische Universit\"at
M\"unchen, 85748 Garching, Germany}
\author{R. Winkler}
\affiliation{Institut f\"ur Festk\"orperphysik, Universit\"at
Hannover, Appelstrasse 2, 30167 Hannover, Germany}
\author{D. Schuh}
\affiliation{Walter Schottky Institut, Technische Universit\"at
M\"unchen, 85748 Garching, Germany}
\author{M. Bichler}
\affiliation{Walter Schottky Institut, Technische Universit\"at
M\"unchen, 85748 Garching, Germany}
\author{M. Grayson}
\affiliation{Walter Schottky Institut, Technische Universit\"at
M\"unchen, 85748 Garching, Germany}

\date{\today}

\begin{abstract}
We measure an anomalous magnetoresistance peak within the lowest Landau level ($\nu = 1$) minimum of a two-dimensional hole system on (110) GaAs. Self-consistent calculations of the valence band mixing show that the two lowest spin-index Landau levels anticross in a perpendicular magnetic field $B$ consistent with where the experimental peak is measured, $B_{\rm p}$. The temperature dependence of the anomalous peak height is interpreted as an activated behavior across this anticrossing gap. Calculations of the spin polarization in the lowest Landau levels predict a rapid switch from about $-3/2$ to $+3/2$ spin at the anticrossing. The peak position $B_{\rm p}$ is shown to be affected by the confinement electrostatics, and the utility of a tunable anticrossing position for spintronics applications is discussed.
\end{abstract}

\pacs{71.70.Di, 71.70.Ej, 72.25.Dc}
\maketitle

High-mobility two-dimensional (2D) systems can be quantized by a perpendicular magnetic field $B$ into Landau levels (LLs), whose discrete energy spectrum is revealed in magnetotransport as the quantum Hall (QH) effect. Anomalous peaks residing within QH minima of the longitudinal resistance provide evidence for crossings or anticrossings of LLs. In AlAs/AlGaAs \cite{poo00}, InP/GaInP \cite{koc93}, Si/SiGe \cite{zei01}, and InGaAs/InAlAs \cite{des04} 2D electron systems, (anti)crossings can be induced in a tilted $B$ field between Landau energies $E_{n,\sigma}= (n+1/2) \hbar \omega_c + \sigma g^\ast \mu_\mathrm{B} B$ with integer $n \ge 0$ and $\sigma = \pm 1/2$ the orbital and spin quantum numbers, $g^\ast$ the effective Land\'e $g$ factor, and $\mu_\mathrm{B}$ the Bohr magneton. The cyclotron frequency $\omega_c = e B_\perp / m^\ast$ depends only on the normal component $B_\perp$ of the total field $B$ whereas the in-plane component $B_\|$ can independently increase the Zeeman energy and induce an (anti)crossing $E_{n,\sigma} \simeq E_{n',\sigma'}$ between levels of differing quantum numbers $n \neq n'$ and $\sigma \neq \sigma'$. In the GaAs valence band, the heavy hole (HH, spin $z$ component $\pm 3/2$) and light hole (LH, $\pm 1/2$) states couple strongly so that the energies of these hole LLs depend highly nonlinearly on $B$ \cite{bro85, eke85}. Crossings or anticrossings of the LLs therefore arise even when the $B$ field is {\it entirely perpendicular} \cite{win03, anticross}. To date, experimental evidence for a hole-LL crossing has been limited to photoluminescence \cite{kub03} and cyclotron resonance \cite{sch85b,sch86,haw93,hir93,col97a} experiments on the (001) facet, and quantatitative agreement with theory has been elusive.
\begin{figure}[!h]
\center \includegraphics[width=9cm]{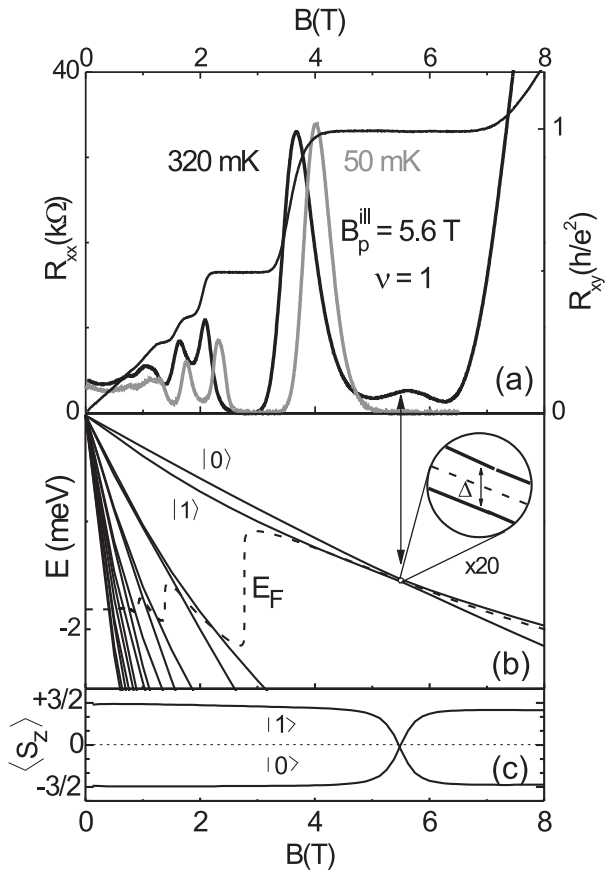}
\caption{(a) Longitudinal $R_{xx}$ and transverse $R_{xy}$ resistance at $T=320$~mK ($50$~mK) shown in black (grey) for the 2D hole system with $p^\mathrm{ill} = 1.34 \times 10^{11}$~cm$^{-2}~(1.40 \times 10^{11}$~cm$^{-2} )$. The small peak in the $\nu = 1$ minimum disappears in the low temperature trace. (b) Self-consistent calculation of the LLs for the system in (a). The dashed line shows the oscillating Fermi energy $E_F$ at $T = 320$~mK. The anticrossing of the two lowest LLs coincides with the peak within the $\nu = 1$ minimum. (c) The average spin $\langle S_z
\rangle$ in the lowest Landau levels, showing a spin flip at the anticrossing.}
\label{fig:exp_theo}
\end{figure}

In this Letter, we report an anomalous resistance peak within the filling factor $\nu = 1$ minimum of the longitudinal resistance of a 2D hole system in (110) oriented GaAs. We observe an activated temperature dependence of the peak height and interpret this as evidence for an anticrossing gap.  Our calculations of the mixing of valence band LLs on the (110) facet in a self-consistent Hartree potential predict such an anticrossing feature, and the $B$ field of the anticrossing coincides with the position $B_{\rm p}$ of the observed peak under the assumption of dilute background doping $\sim 10^{15}~\mathrm{cm}^{-3}$.  The calculated anticrossing shifts with the confinement electric field consistent as observed in the experiment, and spin in the ground state is predicted to switch rapidly with $B$ from $-3/2$ to $+3/2$ in the neighborhood of the anticrossing providing potential applications for spintronics.

The 2D-hole system under study is a modulation doped single interface heterostructure on (110) oriented GaAs using Si as an acceptor. The details of the structure and the growth are reported elsewhere \cite{fis05}. Four-terminal transport measurements are performed in a Hall bar geometry at ac-frequencies between $10 - 30$~Hz and excitation currents between $10 - 200$~nA.  The density is tuned continuously with a thermally evaporated Al-gate on top of the Hall bar. The density can also be changed with persistent photoconductivity via back-side illumination with an infra-red LED. The maximum mobility in this gated sample is $\mu = 125~000$~cm$^2$/Vs at a density of $p = 2.4 \times 10^{11}$~cm$^{-2}$ \cite{sample_name}.

\begin{figure}[t]
\center \includegraphics[width=9cm]{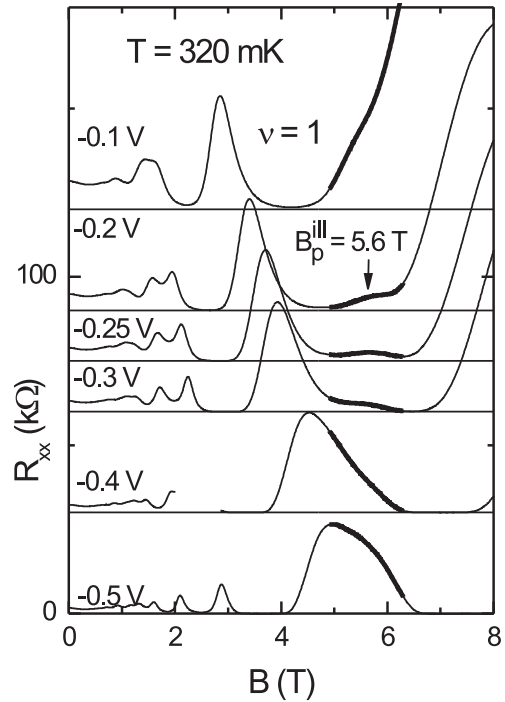}
\caption{$R_{xx}$ for gate voltages $-0.1$~V${} \le V_g \le -0.5$~V corresponding to densities $1.07\times10^{11}$~cm$^{-2} < p < 1.78 \times 10^{11}$~cm$^{-2}$. The anomalous peak is always centered around $B = 5.6$~T as shown with the bold lines. Curves are shifted upward by 30~k$\Omega$ per $0.1$~V gate voltage.}
\label{fig:waterfall}
\end{figure}

Above about 300 mK an anomalous peak appears within the $\nu = 1$ minimum of the longitudinal resistance, whose position depends on the illumination history of the sample but not on the front-gate voltage.  Superscripts will distinguish the two principal illumination states presented here, the dark state (dk) and post-illumination state (ill). Figure \ref{fig:exp_theo}(a) shows the longitudinal $R_{xx}$ and transverse $R_{xy}$ resistance at 320~mK after illumination, with the density $p^\mathrm{ill} = 1.34 \times 10^{11}$~cm$^{-2}$ chosen such that the anomalous peak sits in the center of the $\nu = 1$ minimum.  The peak position $B_\mathrm{p}^\mathrm{ill} = 5.6$~T is reproducible in subsequent cooldowns and does not show any hysteresis, and the peak completely disappears in the 50~mK low temperature trace. Figure \ref{fig:waterfall} shows the weak dependence of the peak position on front-gate bias.  The front-gate voltage tunes the density from from $p = 1.07$ to $1.78 \times 10^{11}$~cm$^{-2}$ shifting $R_{xx}$ features strongly to the right yet leaving the peak position $B_{\rm p}^\mathrm{ill} = 5.6$~T practically unaffected (bold lines).
\begin{figure}[t]
\center \includegraphics[width=9cm]{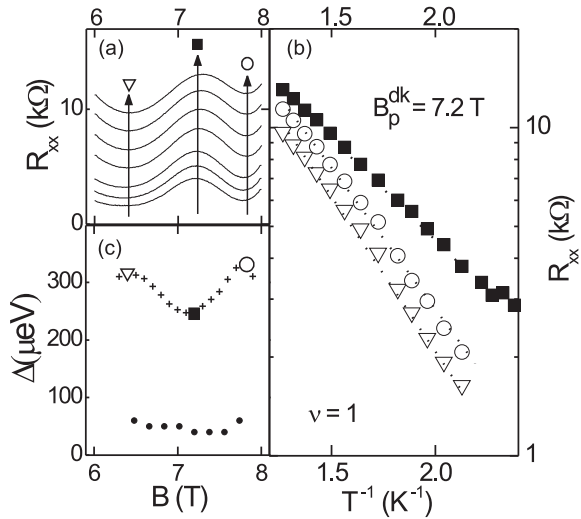}
\caption{(a) Temperature dependence ($400~\textnormal{mK}<T<800~\textnormal{mK}$) of the anomalous peak within the $\nu = 1$ minimum of $R_{xx}$ before illumination with arrows indicating increasing $T$ ($B_{\rm p}^\mathrm{dk} = 7.2$ T, $p^\mathrm{dk} = 1.74 \times 10^{11}~{\rm cm}^{-2}$).  (b) Arrhenius plot of the temperature dependent $R_{xx}$ at the anomalous peak and at the neighboring minima.  (c) The experimental activation gap (\textbf{+}) and calculated anticrossing gap $(\bullet)$ plotted as a continuous function of $B$.}
\label{fig:arrhenius}
\end{figure}%
In contrast, the illumination state of the sample seems to strongly influence the peak position.  In the dark before illumination Fig.~\ref{fig:arrhenius} shows how the peak occurs at $B_{\rm p}^\mathrm{dk} = 7.2$~T for $p^\mathrm{dk} = 1.74 \times 10^{11}~{\rm cm}^{-2}$.  Note that an equivalent density to Fig.~\ref{fig:arrhenius} is reached in Fig.~\ref{fig:waterfall} after illumination, demonstrating that the illumination and not density is the critical factor in shifting the peak position to lower fields.

The temperature dependence of the dark-state $R_{xx}$ within the $\nu~=~1$ QH minimum is shown in Fig.~\ref{fig:arrhenius}(a). The arrows indicate increasing temperature in the range from 400 to 800~mK. If the observed temperature dependence is interpreted in terms of activated conduction across an energy gap, Arrhenius plots such as in Fig.~\ref{fig:arrhenius}(b) can map out the dependence of the gap energy $\Delta$ on $B$ in the neighborhood of the peak, where:
\begin{equation}
R_{xx} \propto \exp (- \Delta /2k_\mathrm{B} T) \,.
\label{eq:arrhenius}
\end{equation}
Arrhenius plots for the $R_{xx}$ peak maximum ($\Box$) and neighboring minima ($\bigtriangledown$, \setlength{\unitlength}{1cm} \put(.2,.07){\circle{0.2}}\hspace{1em}) are shown in Fig.~\ref{fig:arrhenius}(b), with the existence of a local activation minimum consistent with an anticrossing at the resistance peak $\Delta^\mathrm{dk} = 250~\mu$eV in Fig.~\ref{fig:arrhenius}(c).  The same analysis for the illuminated condition results in a gap of $\Delta^\mathrm{ill} = 105~\mu$eV.  At fields further from the anticrossing, the activation gap narrows again presumably because contributions from extended state conduction render Eq.\ (\ref{eq:arrhenius}) invalid.

\begin{figure}[t]
\center \includegraphics[width=9.5cm]{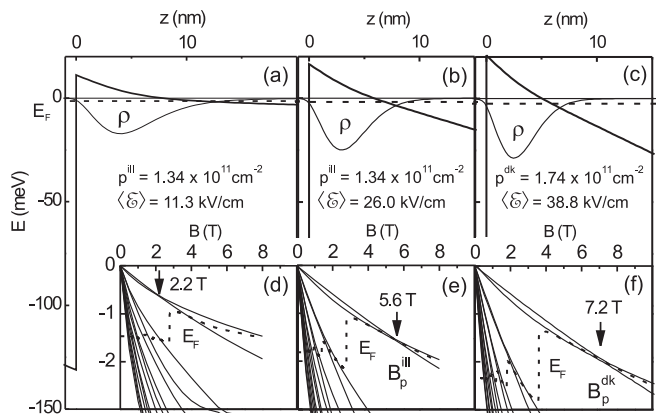}
\caption{Hartree potentials and charge density $\rho(z)$ for a 2DHG with density $p^\mathrm{ill} = 1.34 \times 10^{11}$~cm$^{-2}$ assuming (a) no substrate electric field, and (b) a substrate field which produces a 5.6~T anticrossing. The substrate field in (c) produces an anticrossing at 7.2~T for the density $p^\mathrm{dk} = 1.74 \times 10^{11}$~cm$^{-2}$.  In the corresponding Landau fans (d-f), the ground energy of the 2D subband is defined as $E = 0$. (Fig.~\ref{fig:pot}(e) is reproduced in Fig.~\ref{fig:exp_theo}(b).)}
\label{fig:pot}
\end{figure}

To understand these experimental results, we performed numerical calculations of the 2D hole system based on the methods discussed in Ref.~\cite{win03}.  We first determine the Hartree potential self-consistently at $B=0$ [Fig.~\ref{fig:pot}(a--c)], then we obtain the LLs as a function of $B$ [Fig.~\ref{fig:pot}(d--f)]. For these calculations we use a multiband Hamiltonian that contains the bands $\Gamma_6^c$, $\Gamma_8^v$, and $\Gamma_7^v$, and we take into account all terms of cubic and tetrahedral symmetry which are characteristic of the (110) growth facet of our samples. The hole densities in the calculation correspond either to the gated dark density $p^\mathrm{dk}$ of Fig.~\ref{fig:arrhenius} or to the gated post-illumination density $p^\mathrm{ill}$ of Fig.~\ref{fig:exp_theo}.  In Fig.~\ref{fig:pot}(d), one sees that the uppermost hole LLs do in fact anticross even though $B$ is entirely perpendicular.  This confirms that anticrossings between the lowest LLs occur on this facet and might explain the anomalous $R_{xx}$ peak at $\nu = 1$.

The gate and illumination dependences of the peak position can be explained if one assumes a dilute space charge in the substrate. Note that in the hole system studied here the anticrossing field is determined by the interplay of HH-LH coupling, spin-orbit coupling (Rashba and Dresselhaus term), and the Zeeman term. The former two effects depend on the average normal electric field $\langle {\cal E} \rangle$ seen by the 2D holes in the triangular potential
\cite{win03}
\begin{equation}
\langle {\cal E} \rangle = \frac{e}{\epsilon_r \epsilon_0} \left(
\frac{p} {2} \right) + {\cal E}_\mathrm{S}.
\label{Efield}
\end{equation}
where ${\cal E}_\mathrm{S}$ is the substrate electric field. Figures \ref{fig:pot}(a) and (d) correspond to the case ${\cal E}_\mathrm{S} = 0$. In $p$-type (110) growth, however, persistent photoconductivity \cite{fis05} has been attributed to an illumination-dependent charge state in the buffer layer, whose positive space charge comes from a depleted $n$-type background doping $N_\mathrm{D} \sim 10^{15}~{\rm cm}^{-3}$ \cite{sch85a}. This results in an illumination-dependent substrate electric field ${\cal E}_\mathrm{S} = (e/\epsilon_r \epsilon_0) \int_{\langle z \rangle}^\ell N_\mathrm{D} \, dz$ seen by the 2D holes, where $\langle z \rangle$ is the center of the hole wave function and $\ell$ is the thickness of the space charge layer. The anticrossing calculations can quantitatively account for the observed peak positions at $B_p^\mathrm{ill} = 5.6$ T and $B_p^\mathrm{dk} = 7.2$ T if we treat ${\cal E}_\mathrm{S}$ as a fit parameter.  By choosing ${\cal E}_\mathrm{S}^\mathrm{ill} = 14.7$~V/cm (which would correspond to a uniform $N_\mathrm{D}^\mathrm{ill} = 6.0 \times 10^{14}$~cm$^{-3}$) in Figs.~\ref{fig:pot}(b) and (e), the calculated anticrossing field is matched to the experimentally observed peak position after illumination, $B_\mathrm{p}^\mathrm{ill}$.  Similarly with ${\cal E}_\mathrm{S}^\mathrm{dk} = 24.1$~V/cm ($N_\mathrm{D}^\mathrm{dk} = 1.6 \times 10^{15}$~cm$^{-3}$) the calculated anticrossing aligns with the dark peak position $B_p^\mathrm{dk}$ as in Figs.~\ref{fig:pot}(c) and (f). The resulting $\langle {\cal E} \rangle$ values from Eq.\ (\ref{Efield}) are listed in the figure, and Fig.~\ref{fig:pot}(e) is reproduced in the summary Figure \ref{fig:exp_theo}(b).  $\cal{E}_\mathrm{S}^\mathrm{ill}$ is smaller than $\cal{E}_\mathrm{S}^\mathrm{dk}$, consistent with the persistent photoconductivity observed in these samples.

To observe an anticrossing in transport it is necessary that the anticrossing coincides with the Fermi energy $E_F$ [dashed lines in Fig.~\ref{fig:pot}(d--f)].  The confinement potentials and densities in Figs.~\ref{fig:pot}(b) and (c) are therefore appropriate for observing an anticrossing in the $\nu = 1$ minimum of $R_{xx}$, whereas in Fig.~\ref{fig:pot}(a) no anticrossing feature would be observed.

The calculations underestimate the anticrossing gap measured in experiment.  The size of the calculated gap is 21~$\mu$V for Fig.~\ref{fig:pot}(e) and 41~$\mu$V for Fig.~\ref{fig:pot}(f), about a factor of 5 smaller than the activated gap measurements $\Delta^\mathrm{ill}$ and $\Delta^\mathrm{dk}$, respectively.  It is nonetheless noteworthy that both the calculations and the experiment show a gap which decreases by about a factor of 2 after illumination.  We remark that exchange interactions have been ignored in the calculations and might be partially responsible for larger experimentally observed gap.

Equation (\ref{Efield}) also explains why the position of the anticrossing depends only weakly on the density $p$ of the 2D-hole system tuned by means of a front gate.  In Fig.~\ref{fig:waterfall}, $p$ increases by 67\% producing large shifts in the quantum Hall features.  Yet the average field $\langle {\cal E} \rangle$ seen by the holes only changes by 22\% since the substrate field ${\cal E}_\mathrm{S}$ accounts for the majority of the field. In agreement with these qualitative arguments, our calculations predict that $B_p$ changes by only 0.2~T when $p$ is varied by means of a front gate as in Fig.~\ref{fig:waterfall}. Consistent with these results, careful inspection of Fig.~\ref{fig:waterfall} shows that the peak position indeed shifts slightly to the right with increasing density as a result of the small increase in $\langle {\cal E} \rangle$.

Our calculations can determine the spin polarization of the system, and predict a spin flip at the anticrossing. Figure~\ref{fig:exp_theo}(c) shows the expectation value of the $z$ component of the spin operator, $\langle S_z \rangle$, for the two lowest LLs. Away from the anticrossing, these LLs are almost pure HH LLs with average spin $\langle S_z \rangle$ close to $+3/2$ or $-3/2$.  At the anticrossing, the spin expectations $\langle S_z \rangle$ of these two levels flip from $\pm 3/2$ to $\mp 3/2$ over a small $\sim 1$~T range.

Novel spintronic devices might be able to utilize the electrostatically tunable anticrossing field to control the ground state spin.  One of the most challenging tasks in spintronics is achieving efficient spin injection from ferromagnetic spin-polarized contacts.  According to the present work, a 2D hole sample with patterned front-gates and back-gates could create regions of differing electrostatic confinement and therefore different anticrossing fields $B_{\rm p}$. Assuming $B_{\rm p1} = 5.6$~T and $B_{\rm p2} = 7.2$~T as realized in this experiment, a bias field of $B = (B_{\rm p1} +B_{\rm p2}) / 2 = 6.4$~T applied uniformly over the sample would leave the $B_{\rm p1}$ regions occupied with $S_z \simeq -3/2$ HHs and the $B_{\rm p2}$ regions with $S_z \simeq +3/2$ HHs. Such reservoirs could be used as spin-reservoirs for spintronics and spin-based quantum computations in a reduced two-component pseudo-spin basis.

In summary, we observed an anomalous peak in the longitudinal resistance of two-dimensional hole systems on (110) oriented GaAs. We performed numerical calculations of the Landau levels and found good agreement of the calculated lowest level anticrossing with the experimental peak position. We measured the activation energy of the anticrossing energy gap, and propose that the spin-flip which occurs at the anticrossing will prove useful in spintronics and in spin-based quantum computation.

\begin{acknowledgments}
This work was supported financially by Deutsche For\-schungs\-ge\-mein\-schaft via Schwerpunktprogramm Quantum-Hall-Systeme and in the framework of the COLLECT EC-Research Training Network HPRN-CT-2002-00291. The authors are grateful to K.~Neumaier for assisting with low-temperature measurements.
\end{acknowledgments}


\begin{thebibliography}{10}
\bibitem{anticross}
We get a crossing (an anticrossing) of two LLs $|\alpha\rangle$ and
  $|\beta\rangle$ when these levels transform according to different (the same)
  irreducible double group representations (IDGR) of the point group of the
  system. For 2D systems on a (001) facet the point group at $B_\perp > 0$ is
  $C_2$ with two IDGRs $\Gamma_3$ and $\Gamma_4$ so that we can have both
  crossings and anticrossings. For the low-symmetry facet (110) we get the
  trivial point group $C_1$ with only one IDGR $\Gamma_2$ so that in this case
  we always have anticrossings \cite{win03}. Therefore, we cannot label the LLs
  by spin and orbital quantum numbers, but only by
  one combined index $i=0,1,2,\ldots$. The coupling between two LLs
  $|\alpha\rangle$ and $|\beta\rangle$ can be very weak, however, so that the
  energy gap at an anticrossing is very small, as it is the case in
  Fig.~\ref{fig:exp_theo}. Almost-good quantum numbers are discussed in
  Ref.~\onlinecite{win03}.
\end{thebibliography}
\end{document}